\documentclass[conference, a4paper]{IEEEtran}
\IEEEoverridecommandlockouts

\usepackage{cite}
\usepackage{amsmath,amssymb,amsfonts}
\usepackage{algorithmic}
\usepackage{algorithm}
\usepackage{graphicx}
\usepackage{textcomp}
\usepackage{xcolor}
\usepackage{multirow}
\def\BibTeX{{\rm B\kern-.05em{\sc i\kern-.025em b}\kern-.08em
    T\kern-.1667em\lower.7ex\hbox{E}\kern-.125emX}}
    
\begin{document}

\title{Constructing Knowledge Map for MIMO-OFDM Clustered Channel Estimation \\
\thanks{\copyright  2026 IEEE.  Personal use of this material is permitted.  Permission from IEEE must be obtained for all other uses, in any current or future media, including reprinting/republishing this material for advertising or promotional purposes, creating new collective works, for resale or redistribution to servers or lists, or reuse of any copyrighted component of this work in other works.}
}

\author{
    \IEEEauthorblockN{Heling Zhang\IEEEauthorrefmark{1}\IEEEauthorrefmark{3} Xiujun Zhang\IEEEauthorrefmark{2} Xiaofeng Zhong\IEEEauthorrefmark{1}\IEEEauthorrefmark{3} Shidong Zhou\IEEEauthorrefmark{1}\IEEEauthorrefmark{3}}
    \IEEEauthorblockA{\IEEEauthorrefmark{1}Department of Electronic Engineering, Tsinghua University, Beijing, China}
    \IEEEauthorblockA{\IEEEauthorrefmark{2}Beijing National Research Center for Information Science and Technology}
    \IEEEauthorblockA{\IEEEauthorrefmark{3}State Key Laboratory of Space Network and Communications\\
    Emails: zhanghl24@mails.tsinghua.edu.cn, \{zhangxiujun, zhongxf, zhousd\}@tsinghua.edu.cn}
}

\maketitle
\begin{abstract}
Channel knowledge map (CKM) exploits environment information to assist channel estimation during communication. For clustered channels, which represent a typical type of wireless propagation environment, there has been no research devoted to designing an appropriate CKM to enhance their estimation. To exploit environment information for clustered channel, improve channel estimation accuracy and reduce pilot overhead, we propose ClusterCKM, a CKM providing the range of clustered multipath parameters for any pair of transmitter-receiver links in the region of interest. Firstly, we construct ClusterCKM through estimating the spatial range of scatterer clusters from historical channel information. From these spatial range of scatterer clusters, ClusterCKM infers the range of multipath parameters for the target link. Furthermore, a ClusterCKM-based channel estimation algorithm is developed to utilize the parameter range provided by ClusterCKM. Simulation results show that, more accurate channel estimation can be achieved and pilot overhead can also be reduced by ClusterCKM and the ClusterCKM-based estimation algorithm.
\end{abstract}


\section{Introduction}\label{Sec:Introduction}
Future wireless communication is expected to make a breakthrough on multiple performance indicators, such as latency, peak data rates and connection density\cite{Latvaaho19}. To improve the ability of communication, a precise channel state information is always needed. However, as the scaling of the communication system, the pilot overhead of conventional channel estimation methods like least squares is also increasing sharply, which becomes an unbearable burden for its radio resource consumption.

Channel knowledge map brings hope to reduce pilot overhead by exploiting prior information in the propagation environment. Taking the position of transmitter (Tx) and receiver (Rx) as input, CKM provides environment information for the particular Tx-Rx link to empower communication tasks, such as channel estimation and beamforming\cite{Wu21}. CKM-based communication algorithms may also be tailored to make better use of the output of CKM.

For channel composed of clustered multipaths, the channel is sparse in the joint space-time domain. Providing the multipath parameter range of each cluster through CKM can benefit channel estimation. By exploiting these parameter ranges, the sparsity of the channel can be leveraged to reduce the number of unknowns to be estimated. While some previous studies assume that the entire channel consists of only a few sparse multipath components\cite{Du24,Mundlamuri23}, with the corresponding scatterer positions representing all the environmental information required for channel estimation, clustered channels typically contain dense multipath components generated by clustered scatterers or diffuse reflections\cite{Jiang22}. These dense components often account for a considerable portion of the total channel energy\cite{Richter06,Kotterman05} and cannot be effectively characterized by a few sparse paths\cite{Jiang22}. Nevertheless, their parameters remain bounded by the delay and angular spreads of the clusters\cite{Poutanen11}. Therefore, the parameter range of the multipath components within each cluster represents the intrinsic characteristics of the propagation environment.

Existing studies on clustered channels such as \cite{Xu24} and \cite{Wen21} focus on acquiring the parameters of multipath components within each cluster for sensing and positioning, without considering to utilize these parameters in channel estimation; while  \cite{Wu22} estimates the clustered channel with the assumption that each cluster consists of only several paths, and thus cannot handle dense multipath components. 

Our study proposes ClusterCKM to obtain the parameter ranges of the multipath components within each cluster rather than to estimate multipath parameters, thereby improving the overall estimation performance for clustered channels. Our main contributions are summarized as follows:
\begin{itemize}
    \item We propose ClusterCKM, which provides a multipath parameter range estimation of each channel cluster for any Tx-Rx link;
    \item We present an algorithm to construct such a CKM from historical channel information by locating equivalent scatterers (see section \ref{Sec:CKM Construction}) for each cluster;
    \item We design a channel estimation algorithm based on the output parameter ranges of ClusterCKM, which reduces pilot overhead and improves estimation accuracy.
\end{itemize}


\emph{Notation:} $(\cdot)^\mathbf{H},\ (\cdot)^\mathbf{T},\ (\cdot)^* $ and $(\cdot)^\dagger$ represent the matrix operations of conjugate transpose, transpose, conjugate and pseudo-inverse. Lowercase bold letter $\mathbf{h}$, uppercase bold letter $\mathbf{H}$ and uppercase script letter $\mathcal{H}$ mean a vector, a matrix and a tensor respectively. $\circ_{N}$ stands for the tensor-matrix product on the $N$th dimension, and $\mathcal{H}_{(N)}$ means n-dim tensor unfolding.

\section{System Model and Problem Formulation}\label{Sec:System Model and Problem Formulation}
We consider a ClusterCKM-based channel estimation system which consists of three components: channel, pilot sending, ClusterCKM. These components are modeled in sequence, and finally we formulate the ClusterCKM construction and channel estimation problem.

\subsection{Channel Model}
\begin{figure}[htbp]
	\centerline{\includegraphics[scale=0.15]{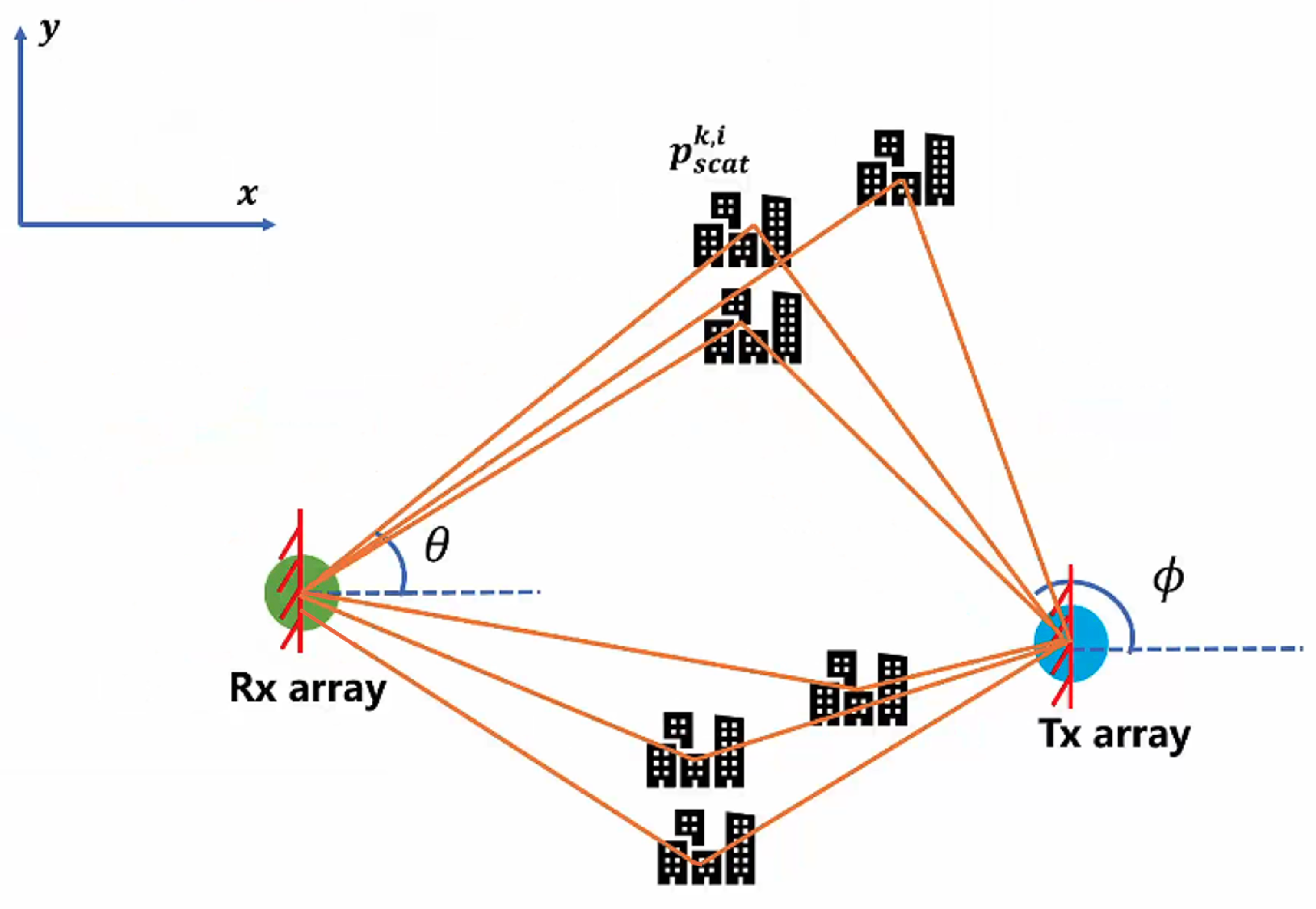}}
	\caption{Considered communication scenario.}
	\label{figure:scheme}
\end{figure}

There are $K$ clusters of scatterers in the environment, and clusters are pairwise disjoint in space. The $k$th cluster contains $L_k$ densely distributed and indistinguishable scatterers. The environment scatterers scatter incident electromagnetic wave to all directions, and have line-of-sight path to all Txs and Rxs in the area. The scattering gain of the $i$th scatterer in the $k$th cluster is $\alpha^{k,i} \in \mathbb{C}$, which integrates the effects of path loss, the structure and the shape of the scatterer. We assume no obstruction exists in the propagation path, so every scatterer in the environment contributes a single path for every Tx-Rx link. Both Tx and Rx in the scenario employ uniform linear array (ULA). The number of antenna elements are $N_{Tx}$ and $N_{Rx}\ (N_{Rx} \gg N_{Tx})$, and the antenna intervals are $d_{Tx}$ and $d_{Rx}$ respectively. The distance from Tx and Rx to each scatterer is far enough and the scattered electromagnetic wave can be regarded as planar wave at the transceiver sites. The $N_{sc}$ subcarriers of the OFDM are uniformly spaced with an interval of $\Delta f$, and the carrier wavelength is $\lambda$. Ignoring the line-of-sight path, which is irrelevant to environment clusters, and multiple scattering, the baseband MIMO-OFDM channel can be organized as a tensor
\begin{equation}\label{eq1}
\begin{aligned}
\mathcal{H} = &\sum_{k=1}^K \sum_{i=1}^{L_k} \alpha^{k,i} \circ_1 \boldsymbol{\gamma}_{N_{Rx}}(\frac{d_{Rx}}{\lambda}sin\theta^{k,i})\\&\circ_2\boldsymbol{\gamma}_{N_{Tx}}(\frac{d_{Tx}}{\lambda}sin\phi^{k,i}) \circ_3 \boldsymbol{\gamma}_{N_{sc}}(\Delta f\tau^{k,i}),
\end{aligned}
\end{equation}
where $\boldsymbol{\gamma}_N(\omega)$ is a sequence of length $N$ and frequency $\omega$
\begin{equation}\label{eq2}
\boldsymbol{\gamma}_N(\omega) := [1, e^{-j2\pi \omega},...,e^{-j2\pi \omega (N-1)}].
\end{equation}
The element $\mathcal{H}(m,n,l)$ corresponds to the channel gain on the $l$th subcarrier from the $n$th Tx antenna to the $m$th Rx antenna. $\theta^{k,i}, \phi^{k,i}, \tau^{k,i}$ respectively stand for the angle of arrival (AoA), the angle of departure (AoD) and the time delay of the path generated by the scattering of the $i$th scatterer in the $k$th cluster located at $\mathbf{p}_{scat}^{k,i}$. The positions of Tx, Rx are denoted as $\mathbf{p}_{Tx}, \mathbf{p}_{Rx}$. Fig. \ref{figure:scheme} illustrates the communication scenario of the channel model.

\subsection{Pilot Sending}
For channel estimation, pilots are transmitted in the frequency domain periodically
\begin{equation}
\mathcal{Y} = \bar{\mathcal{H}} \circ_2 \mathbf{X}^\mathbf{T} + \mathcal{N},
\end{equation}
where the same pilot sequence $\mathbf{X}\in \mathbb{C}^{N_{Tx}\times p}$ is transmitted every $p_f$ subcarriers in the frequency domain, and occupies $p$ OFDM symbols in the time domain. $\bar{\mathcal{H}} \in \mathbb{C}^{N_{Rx}\times N_{Tx}\times \frac{N_{sc}}{p_f}}$ consists of the channel corresponding to the subcarriers carrying pilots, while $\mathcal{Y}$ and $\mathcal{N}$ mean the received signal and the additive Gaussian noise of variance $\sigma_n^2$ respectively.
\subsection{ClusterCKM}

Denote the range of multipath parameters for cluster $k$ as
\begin{equation}\label{eq6}
	\Omega_k = [\theta^k_{min}, \theta^k_{max}] \times  [\phi^k_{min}, \phi^k_{max}] \times  [\tau^k_{min}, \tau^k_{max}],
\end{equation}
ClusterCKM serves as a model which takes the position of Tx and Rx as input, and outputs $K$ estimations of multipath parameter range for all $K$ clusters. The positioning may be inaccurate, but the error is small relative to the distance between Tx/Rx and clusters. Rx has access to several historical Tx-Rx position pairs and the channel information at these positions, so that CKM can be constructed at the Rx side.
\subsection{Problem formulation}
Denoting the historical construction data by the subscript $s$ and the input/output data of CKM by the subscript $c$, we formulate the CKM construction problem as
\begin{itemize}
    \item Given $N_{obs}$ channel information with error and their Tx-Rx position labels $\Theta = \{\hat{\mathcal{H}}_{s,i},\mathbf{p}_{Tx,s,i},\mathbf{p}_{Rx,s,i}\},\ i=1,2,\dots,N_{obs}$, construct a mapping $f_\Theta(\mathbf{p}_{Tx,c}, \mathbf{p}_{Rx,c})$ that maps any given Tx-Rx position pair $\mathbf{p}_{Tx,c}, \mathbf{p}_{Rx,c}$ to $k$ ranges of multipath parameters $\{\hat{\Omega}_k\},\ k=1,2,\dots,K$.
\end{itemize}

Taking outputs of ClusterCKM as prior, we formulate our CKM-based channel estimation problem as
\begin{itemize}
    \item Given  $\{\hat{\Omega}_k\},\ k=1,2,\dots,K$ as environment information and the received signal $\mathcal{Y}$, estimate the MIMO-OFDM channel $\hat{\mathcal{H}}_c$ with known pilot sequence $\mathbf{X}$.
\end{itemize}
\section{ClusterCKM Construction} \label{Sec:CKM Construction}
We firstly divide the historical channel information from $\Theta$ into $\hat{K}$ separated single-cluster channels for each scatterer cluster. From these single-cluster channels, \emph{equivalent scatterers} (ES), which will be defined later in this section, are located by performing tensor decomposition and harmonic analysis on the single-cluster channel tensors. Finally, for any target link with Tx-Rx positions, the range of multipath parameters can be estimated with the help of the estimated positions of ESs.
\subsection{Channel division}
As will be shown in section \ref{Sec:Simulation Results}, if we directly perform tensor decomposition on the channel information $\hat{\mathcal{H}}_s\in \Theta$, the ES positions computed from the result will contain a considerable number of spurious artifacts, with only a few ESs corresponding to each cluster. In this situation, the accuracy of the parameter range estimation will suffer severe degradation. To avoid this problem, we firstly separate $\hat{\mathcal{H}}_s$ into
$\hat{K}$ single-cluster channels $\hat{\mathcal{H}}_s^k, k = 1,\dots,\hat{K}$.
 
With the parameter of each path in a cluster constrained in a limited range, in the domain of AoA, AoD and delay, the main part of a single-cluster channel is respectively located on a low-dimensional subspace. Let the subspaces on these domains be spanned by the column vectors of matrices $\boldsymbol{\Gamma}^k_{s,AoA}$, $ \boldsymbol{\Gamma}^k_{s,AoD}$ and $\boldsymbol{\Gamma}^k_{s,delay}$, then a single-cluster channel tensor has the form of Tucker decomposition
\begin{equation}\label{eq16}
	\hat{\mathcal{H}}^k_s \approx \hat{\mathcal{G}}^k_s \circ_1 \boldsymbol{\Gamma}^k_{s, AoA} \circ_2  \boldsymbol{\Gamma}^k_{s, AoD}\circ_3  \boldsymbol{\Gamma}^k_{s, delay}.
\end{equation}
where $\hat{\mathcal{G}}^k_s$ is a core tensor. With the cluster separation assumption in (\ref{eq6}), the subspaces where different single-cluster channels are located are nearly orthogonal. The $k$th single-cluster channel can thus be separated from the channel information by projection
\begin{equation}\label{eq12}
	\begin{aligned}
		\hat{\mathcal{H}}^k_s  &\approx  \hat{\mathcal{H}}_s \circ_1 \boldsymbol{\Gamma}^k_{s,AoA}\boldsymbol{\Gamma}^{k\ \mathbf{H}}_{s,AoA} \circ_2\\&\boldsymbol{\Gamma}^k_{s,AoD} \boldsymbol{\Gamma}^{k\ \mathbf{H}}_{s,AoD} \circ_3 \boldsymbol{\Gamma}^k_{s,delay} \boldsymbol{\Gamma}^{k\ \mathbf{H}}_{s,delay}.
	\end{aligned}
\end{equation}

Sequentially, we identify $\{\boldsymbol{\Gamma}^k_{s,AoA}, \boldsymbol{\Gamma}^k_{s,AoD},\boldsymbol{\Gamma}^k_{s,delay}\}$ for each cluster according to the descending order of cluster energy. To find proper matrices for the strongest cluster, we firstly perform rank-1 enhanced CP decomposition\cite{Gong22} on channel information $\hat{\mathcal{H}}_s$. Denote component vectors of the CP decomposition result as $\mathbf{a}$, $\mathbf{b}$, $ \mathbf{c}$. These three vectors correspond to the center position of the strongest cluster, and path parameters related to this center can be extracted by performing MUSIC algorithm \cite{Schmidt86} to $\mathbf{a}$, $\mathbf{b}$ and $ \mathbf{c}$
\begin{align}
	\hat{\theta}_{center} &= m_{\theta} (\mathbf{a})\label{eq9},\\
	\hat{\phi}_{center} &= m_{\phi} (\mathbf{b})\label{eq10},\\
	\hat{\tau}_{center} &= m_{\tau}(\mathbf{c})\label{eq11},
\end{align}
where $m(\cdot)$ stands for MUSIC algorithm \cite{Schmidt86}.
 
From the prior information that scatterers are clustered, it can be inferred that the channel power of this strongest cluster is concentrated around these parameters. In the domain of AoA, we construct a set of orthonormal bases
\begin{equation}\label{eq13}
	\bar{\boldsymbol{\gamma}}^{1,i}_{s,AoA} = \frac{1}{\sqrt{N_{Rx}}} \boldsymbol{\gamma}_{N_{Rx}}(\frac{d_{Rx}}{\lambda}\hat{\theta}^1_{center}+\frac{i}{N_{Rx}}),
\end{equation}
and project the 1-dim unfolding of $\hat{\mathcal{H}}_s$ onto them. The power of the channel projection is evaluated as
\begin{equation}\label{eq14}
	p^{1,i}_{s,AoA}=\Vert \bar{\boldsymbol{\gamma}}^{1,i\ \mathbf{H}}_{s,AoA} \hat{\mathcal{H}_s}_{(1)}\Vert_F ^2.
\end{equation}
If the power exceeds a predetermined threshold, the basis $\bar{\boldsymbol{\gamma}}^{1,i}_{s,AoA}$ is reserved in $\boldsymbol{\Gamma}^1_{s,AoA}$. The other two matrices $\boldsymbol{\Gamma}^1_{s,AoD}$ and $\boldsymbol{\Gamma}^1_{s,delay}$ are determined likewise. A larger threshold implies the use of less pilot resource but tolerates higher CKM construction and channel estimation errors, and vice versa.

After obtaining $\{\boldsymbol{\Gamma}^1_{s,AoA},  \boldsymbol{\Gamma}^1_{s,AoD}, \boldsymbol{\Gamma}^1_{s,delay}\}$, we project $\hat{\mathcal{H}}_s$ to the subspace corresponding to the strongest cluster by  (\ref{eq12}) to separate the first single-cluster channel $\hat{\mathcal{H}}^1_s$. Subtract $\hat{\mathcal{H}}^1_s$ from the entire channel $\hat{\mathcal{H}}_s$ and repeat the procedure above until the remaining energy of $\hat{\mathcal{H}}_s$ becomes sufficiently small, all $\hat{K}$ single-cluster channels are separated. Some clusters still exhibit large residuals after being separated according to  (\ref{eq12}); these clusters are extracted multiple times, so $\hat{K}$ may be slightly larger than the real amount of clusters $K$.

\subsection{Equivalent Scatterer Positioning}
If the spatial ranges of scatterers cluster are known, the ranges of multipath parameters can be also derived for an arbitrary Tx-Rx position pair. However, given that the scatterers in the same cluster are densely distributed and indistinguishable, it's impossible to estimate the spatial ranges of clusters by locating every single scatterer from the single-cluster channels. To solve this problem, we introduce the concept of \emph{equivalent path} and \emph{equivalent scatterer} as a substitute for the physical paths and scatterers in the cluster. 

An \emph{equivalent path} (EP) is defined as the approximation to the sum of a group of paths sharing similar parameters. In the considered system which only includes single scattering, these paths are generated by spatially adjacent scatters. An \emph{equivalent scatterer} is defined as the representation of these adjacent scatterers and is located approximately at their center, contributing to the channel with an EP. With the concept of EP, rewrite  (\ref{eq1}) as
\begin{equation}\label{eq8}
\hat{\mathcal{H}}_{s} = \sum_{k=1}^{\hat{K}}\sum_{i=1}^{N_k} g^{k,i}\circ_1 \mathbf{a}^{k,i}\circ_2 \mathbf{b}^{k,i}\circ_3 \mathbf{c}^{k,i} + \epsilon.
\end{equation}
Here the channel from the $k$th scatterer cluster is approximated by $N_k$ EPs, and $\epsilon$ stands for the residual error of CP tensor decomposition. Each EP is a rank-1 tensor and is denoted as $\mathbf{a}^{k,i}\circ_2 \mathbf{b}^{k,i}\circ_3 \mathbf{c}^{k,i}$. 

According to \cite{Wen21}, for a single-cluster channel, though the dense environmental scatterers cannot be recovered, ESs can be located through performing harmonic analysis algorithm on the component vectors $\mathbf{a}^{k,i},\  \mathbf{b}^{k,i},\ \mathbf{c}^{k,i}$ of their corresponding EPs. The spatial range of the obtained ES positions in a cluster is approximately equal to that of the cluster.

To locate all $N_k$ ESs in the $k$th cluster, for all $\hat{\mathcal{H}}_{s}^k$, we firstly perform rank-$N_k$ CP decomposition to extract their EPs, where $N_k$ is determined by a proper rank-determining criterion. By substituting the component vectors of the obtained EPs into (\ref{eq9}) and (\ref{eq11}), we obtain the path parameter $\hat{\tau}^{k,i}_{ep}$ and $\hat{\theta}^{k,i}_{ep}$ for each ES. Given the position label $\mathbf{p}_{Tx,s}, \mathbf{p}_{Rx,s}$ provided with the channel information data $\hat{\mathcal{H}}_s$, $\hat{\tau}^{k,i}_{ep}$ determines an ellipse whose two foci are on the position of Tx and Rx, while $\hat{\theta}^{k,i}_{ep}$ determines a ray starts at the position of Rx. The position of the ES $\hat{\mathbf{p}}^{k,i}_{es}$ is thus estimated by solving the intersection between the ellipse and the ray.

\subsection{Multipath Parameter Range Estimation}
To generate outputs of ClusterCKM, we extract ES positions from all $\hat{\mathcal{H}}_s\in \Theta$ through the procedure above. These positions are then re-clustered into $\tilde{K}$ clusters to merge overlapping clusters. $\tilde{K}$ is determined by Davies-Bouldin index \cite{Davies79}, and outliers outside the clusters are removed. Subsequently, for any input Tx-Rx position, we calculate the parameter for each path connecting Tx, Rx and one of the ESs. The parameter range of all paths in a single cluster is denoted as $\hat{\Omega}_k$, and is taken as the estimation of the real parameter range.
\section{CKM-based Channel Estimation} \label{Sec:Channel Estimation}
In this section, we convert the multipath parameter ranges provided by ClusterCKM to a series of subspace-determining matrices. Channels in the determined subspaces are estimated sequentially, and then interpolated from the subcarriers carrying pilot to all subcarriers. Finally, estimations in all subspaces are summed up to yield the overall estimation $\hat{\mathcal{H}}_c$.

For a given Tx-Rx link in the considered communication scenario, the channel $\bar{\mathcal{H}}_c$ on subcarriers carrying pilot contains $K$ single-cluster channels $\bar{\mathcal{H}}^k_c,\ k=1,\dots,K$. Each of these channels, from (\ref{eq16}), can be approximated by a Tucker decomposition composed of core tensor $\bar{\mathcal{G}}^k_c$ and subspace matrices $\boldsymbol{\Gamma}^k_{c, AoA}$, $\boldsymbol{\Gamma}^k_{c, AoD}$ and $\boldsymbol{\Gamma}^k_{c, delay}$. Error of this Tucker approximation comes from the power leakage outside the subspace determined by these matrices. If they are properly chosen to make the leakage negligible, $\bar{\mathcal{H}}^k_c$ can be estimated through estimating $\bar{\mathcal{G}}^k_c$, which has much fewer elements.

With the estimated multipath parameter range $\hat{\Omega}_k$ known, and assuming that channel power is uniformly distributed in the parameter range, the optimal choice for $\boldsymbol{\Gamma}^k_{c, AoA}$ with minimum power leakage contains a series of discrete prolate spheroidal sequences (DPSS)\cite{Slepian78}. By substituting the digital frequency range $[\frac{d_{Rx}}{\lambda}sin(\hat{\theta}^k_{min}),\frac{d_{Rx}}{\lambda}sin(\hat{\theta}^k_{max})]$ derived from $\hat{\Omega}_k$ into formula from \cite{Slepian78}, we generate $N_{dpss}=\lceil N_{Rx}(\frac{d_{Rx}}{\lambda}sin(\hat{\theta}^k_{max})-\frac{d_{Rx}}{\lambda}sin(\hat{\theta}^k_{min}))\rceil+1$ DPSSs as columns of $\boldsymbol{\Gamma}^k_{c, AoA}$ to minimize power leakage while restricting the element number of $\bar{\mathcal{G}}_c^k$. Similarly, we can also generate $\boldsymbol{\Gamma}^k_{c, AoD}$ and $\boldsymbol{\Gamma}^k_{c, delay}$ from the same procedure.

After obtaining the subspace matrices of the $k$th single-cluster channel, the core tensor $\mathcal{G}^k_c$ of the $k$th cluster subspace is then estimated by
\begin{equation}\label{eq18}
	\hat{\mathcal{G}}^k_c = \mathcal{Y} \circ_1 \boldsymbol{\Gamma}^{k\ \mathbf{H}}_{c, AoA} \circ_2  \boldsymbol{\Gamma}^{k\ \mathbf{H}}_{c, AoD}(\mathbf{X}^\mathbf{T})^\dagger\circ_3  \boldsymbol{\Gamma}^{k\ \mathbf{H}}_{c, delay}.
\end{equation}
$\hat{\mathcal{G}}^k_c$ depends on the output of CKM through the three subspace matrices. Such $\hat{\mathcal{G}}^k_c$, however, cannot be used to recover the single-cluster channel directly. Due to possible error in ES localization, the output of ClusterCKM may not be accurate enough and need to be further optimized. Setting $\hat{\Omega}_k$ as the initial value, the multipath parameter range is  
adjusted in finite steps in the neighborhood of $\hat{\Omega}_k$ by maximizing the power of $\hat{\mathcal{G}}^k_c$, which represents the power of single-cluster channel within the subspace matrices determined by the multipath parameter range. The optimal value is denoted as $\hat{\Omega}_k^*$, and the corresponding core tensor from (\ref{eq18}) is denoted as $\hat{\mathcal{G}}^{k*}_c$.

All three subspace matrices are then determined by the optimal multipath parameter range through generating DPSSs and are denoted as $\boldsymbol{\Gamma}^{k*}_{c,AoA},  \boldsymbol{\Gamma}^{k*}_{c,AoD},  \boldsymbol{\Gamma}^{k*}_{c,delay}$ respectively. The $k$th single-cluster channel is estimated by
\begin{equation}\label{eq19}
\hat{\mathcal{H}}_c^k= \hat{\mathcal{G}}^{k*}_c \circ_1 \boldsymbol{\Gamma}^{k*}_{c, AoA} \circ_2  \boldsymbol{\Gamma}^{k*}_{c, AoD}\circ_3  \boldsymbol{\Gamma}^{k*}_{c, delay}.
\end{equation}
After the first single-cluster channel $\hat{\mathcal{H}}_c^1$ is estimated, the overall received signal is updated by subtracting the contribution of this cluster
\begin{equation}\label{eq20}
    \mathcal{Y} \leftarrow \mathcal{Y}-\hat{\mathcal{H}}_c^1\circ_2 \mathbf{X^T},
\end{equation}
from $\tilde{K}$ parameter ranges from ClusterCKM, $\tilde{K}$ single-cluster channels are estimated through repeating the procedure above.

To interpolate $\hat{\mathcal{H}}_c^k$, which is the channel estimation on the subcarriers carrying pilot, to all subcarriers, we generate another matrix $\tilde{\boldsymbol{\Gamma}}^{k*}_{c, delay}$ to provide a subspace on all $N_{sc}$ subcarriers and substitute $\boldsymbol{\Gamma}^{k*}_{c, delay}$ with it. Given that both $\boldsymbol{\Gamma}^{k*}_{c, delay}$ and $\tilde{\boldsymbol{\Gamma}}^{k*}_{c, delay}$ describe the delay feature of the same cluster, the frequency range and the number for DPSSs to determine $\tilde{\boldsymbol{\Gamma}}^{k*}_{c, delay}$ is the same as $\boldsymbol{\Gamma}^{k*}_{c, delay}$. However, the length of DPSSs for $\tilde{\boldsymbol{\Gamma}}^{k*}_{c, delay}$ is $N_{sc}$ while it is only $\frac{N_{sc}}{p_f}$ for $\boldsymbol{\Gamma}^{k*}_{c, delay}$. The single-cluster channel estimation is then interpolated to all subcarriers in the frequency domain as
\begin{equation}\label{eq21}
    \check{\mathcal{H}}_c^k = \hat{\mathcal{H}}_c^k \circ_3 \tilde{\boldsymbol{\Gamma}}^{k*}_{c, delay} \boldsymbol{\Gamma}^{k*\ \mathbf{H}}_{c, delay},
\end{equation}
and the summation of all single-cluster channels yields the overall estimation $\hat{\mathcal{H}}_c$.


\section{Simulation Results} \label{Sec:Simulation Results}
In this section, we firstly examine the effectiveness of our method for CKM construction. Secondly, we verify the ability of ClusterCKM to improve channel estimation under different pilot periods in frequency domain. Finally, we change the environmental conditions to evaluate the performance of ClusterCKM in different environments.

All compared schemes are listed below.
\begin{itemize}
    \item Least Square (LS): Plain least square channel estimation without knowing any environment prior information. The channel is interpolated with FFT.
    \item OMP: Channel estimation based on orthogonal matching pursuit (OMP)\cite{Pati93}. The dictionary of OMP is set as orthonormal column vectors, each of which is a Kronecker product of steering vectors in the domain of AoA, AoD and time delay. The channel is interpolated with FFT.
    \item ClusterCKM: CKM-based channel estimation in section \ref{Sec:Channel Estimation} with ClusterCKM constructed in section \ref{Sec:CKM Construction}.
    \item CoarseCKM: Same as ClusterCKM, but ESs are located directly from historical channel information without separating the channel into single-cluster channels. 
    \item SparseCKM: The channel estimation algorithm in \cite{Du24} with the stage "coarse parameter estimation" substituted by the sensing algorithm in \cite{Gong22}. It is designed for sparse multipath channels but not for clustered channels.
\end{itemize}
\begin{figure}[htbp]
\centerline{\includegraphics[scale=0.14]{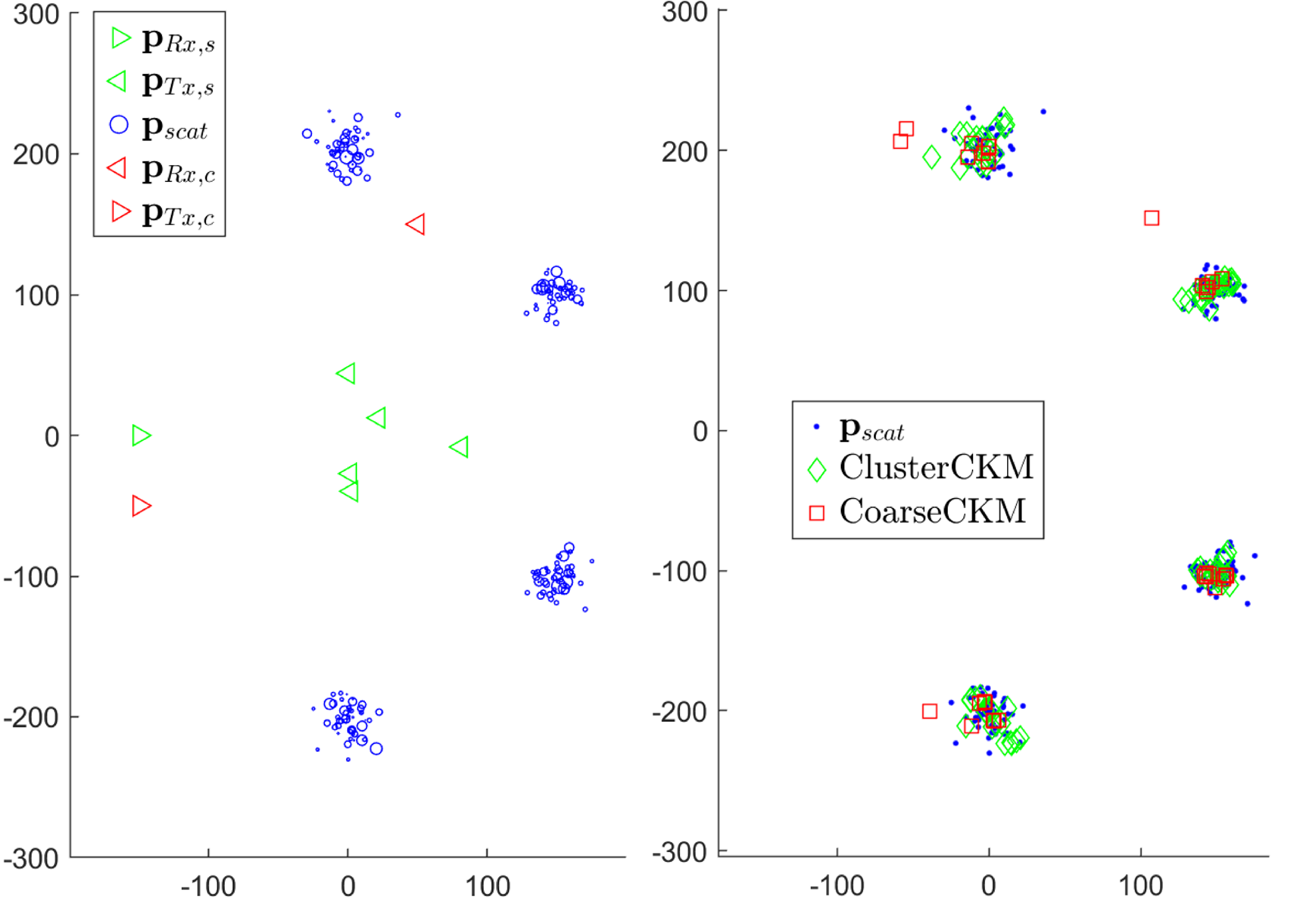}}
\caption{The schematic of the simulation environment (left) and the result of equivalent scatterer localization (right).}
\label{figure:schemetic}
\end{figure}

The simulation environment is shown in the left part of Fig. \ref{figure:schemetic} and the parameters are set as follows. The antenna numbers of Tx and Rx are $N_{Tx}=3$ and $N_{Rx}=16$, and distances between the antenna elements are both $0.5\lambda$. The system bandwidth is BW=100MHz and the number of uniformly distributed subcarriers is 210 for CKM construction and 1680 for communication. $K=4$ clusters exists in the environment, and each cluster includes $L_k=50$ scatterers. The scattering amplitude of each scatterer is randomly generated from a standard complex Gaussian distribution, and decays inversely with the total length of the corresponding path. The center positions of the 4 clusters are $[0,200]^\mathbf{T},\ [0,-200]^\mathbf{T},\ [150,-100]^\mathbf{T},\ [150,100]^\mathbf{T}$. Positions of scatterers are generated randomly from a 2D Gaussian distribution, whose average and variance are the center of the corresponding cluster and $(10m)^2$. The historical channel information $\hat{\mathcal{H}}_s$ is generated at $N_{obs}=5$ randomly chosen Tx positions and a fixed Rx position $\mathbf{p}_{Rx,s}=[-150,0]^\mathbf{T}$ with an error of 10dB compared to the accurate data $\mathcal{H}_s$.  

For channel estimation, pilot $\mathbf{X}$ is a randomly generated row orthogonal matrix with $\Vert \mathbf{X} \Vert_F^2=p$, and the pilot length in time domain is set to be $p=3$. The SNR is defined as $\frac{\Vert\mathcal{H}_c\Vert_F^2}{N_{Tx}N_{Rx}N_{sc}\sigma_n^2}$ and is adjusted during the simulation to create different communication situations. Finally, we define RMSE as $\frac{\Vert \hat{\mathcal{H}}_c-\mathcal{H}_c\Vert_F^2}{\Vert\mathcal{H}_c\Vert_F^2}$, and take the average RMSE of 15 program runs as the metric for channel estimation accuracy. The target positions of channel estimation are $\mathbf{p}_{Tx,c}=[50,150]^\mathbf{T}$ and $\mathbf{p}_{Rx,c}=[-150,50]^\mathbf{T}$.

\subsection{Effectiveness of CKM Construction}
For evaluating the effectiveness of ClusterCKM constructed by our algorithm and proving the necessity of channel separation in CKM construction, we exhibit the localization performance of ESs obtained with and without channel separation (labeled "ClusterCKM" and "CoarseCKM", respectively) on the right side of Fig. \ref{figure:schemetic}. The ES positions produced with channel separation are nearly uniformly distributed within the spatial range of each scatterer cluster, thus guaranteeing the accuracy of multipath parameter range estimation. In contrast, fewer ESs are generated without channel separation, which are insufficient to capture the spatial range of the clusters, and some of the estimated scatterer positions even fall outside the actual cluster regions, resulting in deviations of ClusterCKM constructed from the estimated ES positions.

\begin{figure}[htbp]
\centerline{\includegraphics[scale=0.25]{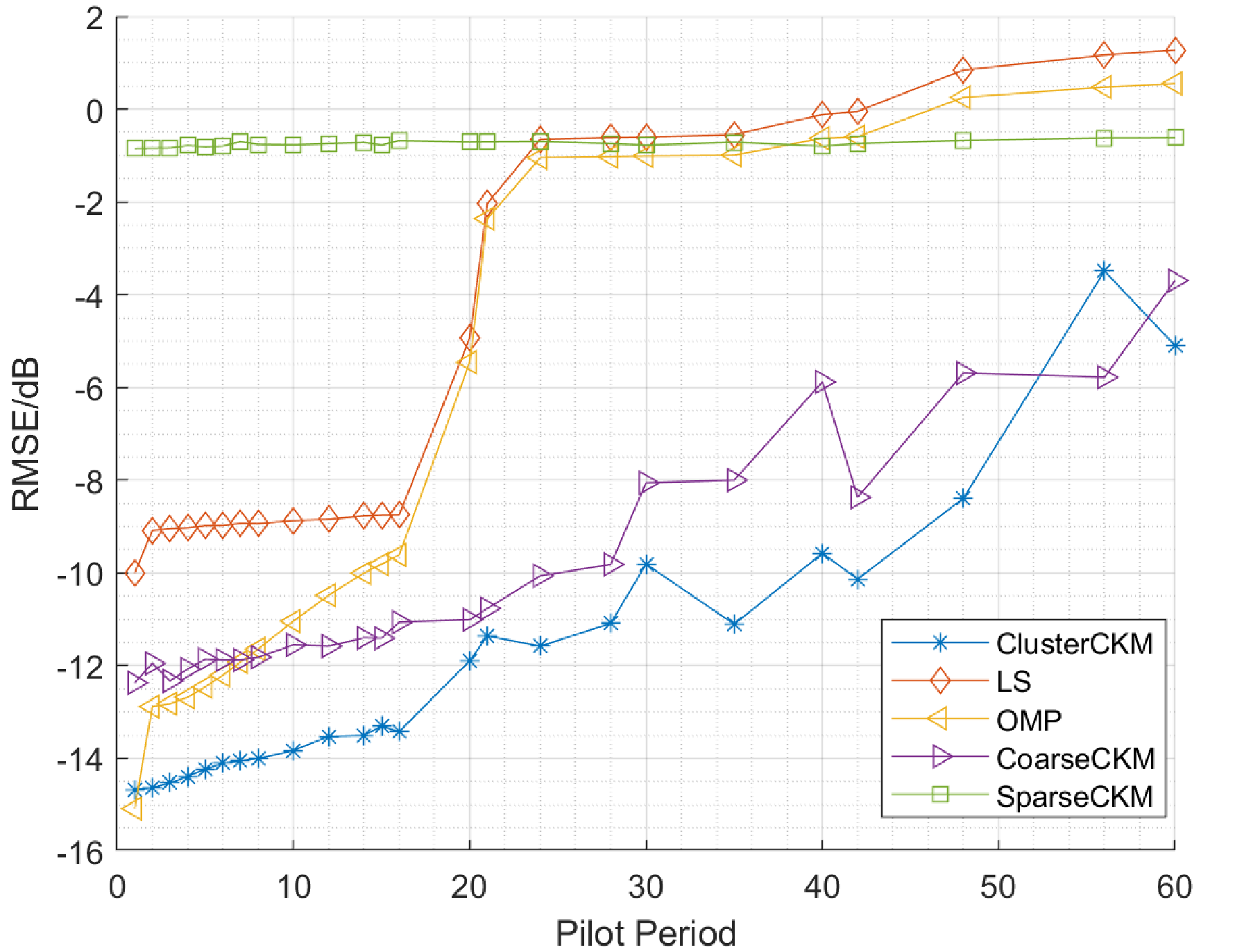}}
\caption{Channel estimation accuracy with different pilot periods.}
\label{figure:experiment1}
\end{figure}

\subsection{Channel Estimation}

\begin{table*}[htbp]
\centering
\caption{Channel estimation accuracy (RMSE) of compared schemes in different environments (in \textnormal{dB})} 
\begin{tabular}{|c|cccc|cccc|}
\hline
\multirow{2}{*}{\shortstack{ \\ \\Estimation\\Scheme}} & \multicolumn{4}{c|}{SNR=20dB} & \multicolumn{4}{c|}{SNR=0dB} \\\cline{2-9}
 & Sparse & 1 large cluster & 4 small clusters & 4 large clusters & Sparse & 1 large cluster & 4 small clusters & 4 large clusters \\
  & $p_f=20$ & $p_f=20$ & $p_f=20$ & $p_f=15$ & $p_f=20$ & $p_f=20$ & $p_f=20$ & $p_f=15$ \\
\hline
ClusterCKM &-13.0&-13.4&\textbf{-15.4}&\textbf{-16.4}&-10.5&\textbf{-7.0}&\textbf{-5.8}&\textbf{-5.8}\\
LS &\ 0.0&\textbf{-14.8}&-6.5&-10.3&\ 0.0&\ 0.8&\ 1.8&\ 0.0\\
OMP &-4.8&-11.6&-5.8&-7.5&-4.8&-4.2&-2.5&-4.2\\
CoarseCKM &-12.0&-14.9&-12.0&-7.0&-11.6&-6.7&-5.1&-5.1\\
SparseCKM &\textbf{-30.7}&-0.4&-0.6&-0.5&\textbf{-27.0}&-0.2&-0.5&-0.4\\
\hline
\end{tabular}
\label{tab:environment}
\end{table*}

We vary the frequency-domain pilot period $p_f$ from 1 to 60 with SNR fixed at 10 dB and compare the performance of different schemes. The result is shown in Fig. \ref{figure:experiment1}.

When the pilot is sufficient, ClusterCKM-based estimation improves estimation accuracy with fine environment information. For $p_f \leq 16$, CoarseCKM, ClusterCKM and OMP all outperform LS, in which the ClusterCKM realizes the highest accuracy. This is because these three schemes all search the subspace where the channel power is concentrated to reduce the number of elements to be estimated. By utilizing the accurate multipath parameter range estimation, ClusterCKM can find the most appropriate subspace, thus improving accuracy.

Furthermore, ClusterCKM reduces pilot consumption by allowing a larger pilot period. When $16 < p_f \leq 48 $, the pilot spacing exceeds the coherence bandwidth, causing LS and OMP to degrade sharply. However, CoarseCKM and ClusterCKM, which adopt the proposed CKM-based algorithm, perform per-cluster estimation. For these schemes, the coherence bandwidth depends only on the maximum delay spread of individual clusters rather than the total channel delay spread, effectively enlarging the coherence bandwidth. With ClusterCKM, the maximum pilot period to achieve an RMSE of -8dB increases to $p_f=48$ from $p_f=16$ for LS and OMP, saving pilot overhead by $2/3$. The fluctuation in accuracy as $p_f$ increases is due to that multipath parameters corresponding to the scatterers are not evenly spaced, making some pilot periods more suitable for channel estimation than others.

For $p_f > 48$, the pilot period is too large for accurate interpolation with any scheme. Specifically, SparseCKM consistently fails due to model mismatch.
\subsection{Performance in different environments}
We continue to evaluate the performance of ClusterCKM in different environments. The results are shown in the Table \ref{tab:environment}, where "Sparse" means each cluster only contains 1 path ($L_k=1$), "small cluster" and "large cluster" mean that all clusters have a radius of 10m and 20m. For "1 large cluster", the cluster center position is $[0,-200]^\mathbf{T}$, and for other environment settings the positions of cluster centers remain the same as above. The SNR and 
$p_f$ settings are given in the table header; all other parameters are also the same as above.

At high SNR, ClusterCKM enables reliable estimation performance under various channel conditions. LS and OMP fail to perform correct interpolation when multiple clusters exist, while CoarseCKM suffers from inaccurate ES localization in the presence of several large clusters. Meanwhile, SparseCKM is applicable only to sparse environments. In contrast, ClusterCKM consistently extracts accurate cluster parameter ranges from historical channel information, leading to precise channel estimation across all scenarios. 

At low SNR, accurate parameter range estimation becomes less critical, but ClusterCKM still provides gain to estimation accuracy. The performance of CoarseCKM remains close to that of ClusterCKM, since the estimation error is mainly caused by additive noise rather than CKM bias. Both CKM-based schemes outperform LS and OMP, benefiting from their ability to interpolate effectively with larger pilot periods.

\section{Conclusion} \label{Sec:Conclusion}
In this paper, we propose ClusterCKM tailored for clustered channels to provide multipath parameter range estimation of each channel cluster for any Tx-Rx position. To construct ClusterCKM, we estimate the positions of equivalent scatterers and compute multipath parameter ranges from the estimations. Furthermore, outputs of ClusterCKM are utilized by our CKM-based channel estimation algorithm, which infers the channel subspace associated with each cluster and estimates the channel in each subspace sequentially. Simulation results demonstrate that the proposed method enables high-quality construction of ClusterCKM using only a small number of historical channel information. Furthermore, leveraging the environment prior information provided by ClusterCKM, our CKM-based channel estimation significantly reduces pilot overhead by $2/3$ compared to conventional methods, while achieving substantial improvements in estimation accuracy and being robust in various environments.


\begin{thebibliography}{00}

\bibitem{Latvaaho19} 6G Research Visions 1: Key Drivers and Research Challenges for 6G Ubiquitous Wireless Intelligence: 6G Flagship, M. Latvaaho, K. Leppanen, Eds. Oulu, Finland: Univ. Oulu, Sep. 2019.
\bibitem{Wu21} D. Wu, Y. Zeng, S. Jin and R. Zhang, "Environment-Aware and Training-Free Beam Alignment for mmWave Massive MIMO via Channel Knowledge Map," 2021 IEEE International Conference on Communications Workshops (ICC Workshops), Montreal, QC, Canada, 2021, pp. 1-7.
\bibitem{Du24} J. Du, Y. Chen, P. Zhang, S. Mumtaz, X. Li and D. B. da Costa, "An Effective Simultaneous Channel Estimation and Sensing Algorithm for mmWave MIMO-OFDM Systems," in IEEE Transactions on Wireless Communications, vol. 23, no. 11, pp. 17054-17069, Nov. 2024.
\bibitem{Mundlamuri23} R. Mundlamuri, R. Gangula, C. K. Thomas, F. Kaltenberger and W. Saad, "Sensing Aided Channel Estimation in Wideband Millimeter-Wave MIMO Systems," 2023 IEEE International Conference on Communications Workshops (ICC Workshops), Rome, Italy, 2023, pp. 1404-1409.
\bibitem{Jiang22} S. Jiang, W. Wang, Y. Miao, W. Fan and A. F. Molisch, "A Survey of Dense Multipath and Its Impact on Wireless Systems," in IEEE Open Journal of Antennas and Propagation, vol. 3, pp. 435-460, 2022.
\bibitem{Richter06} A. Richter, J. Salmi and V. Koivunen, "Distributed scattering in radio channels and its contribution to MIMO channel capacity," 2006 First European Conference on Antennas and Propagation, Nice, France, 2006.
\bibitem{Kotterman05} W. A. T. Kotterman, M. Landmann, G. Sommerkorn, and R. Thomä, “On diffuse and non-resolved multipath components in directional channel characterisation,” in Proc. Gen. Assem. URSI, New Delhi, India, Oct. 2005, pp. 1–4.
\bibitem{Poutanen11} J. Poutanen, J. Salmi, K. Haneda, V.-M. Kolmonen, and P. Vainikainen, “Angular and shadowing characteristics of dense multipath components in indoor radio channels,” IEEE Trans. Antennas Propag., vol. 59, no. 1, pp. 245–253, Jan. 2011.
\bibitem{Xu24} W. Xu, Y. Xiao, A. Liu, M. Lei and M. -J. Zhao, "Joint Scattering Environment Sensing and Channel Estimation Based on Non-Stationary Markov Random Field," in IEEE Transactions on Wireless Communications, vol. 23, no. 5, pp. 3903-3917, May 2024.
\bibitem{Wen21} F. Wen, J. Kulmer, K. Witrisal and H. Wymeersch, "5G Positioning and Mapping With Diffuse Multipath," in IEEE Transactions on Wireless Communications, vol. 20, no. 2, pp. 1164-1174, Feb. 2021.
\bibitem{Wu22} X. Wu, S. Ma, X. Yang and G. Yang, "Clustered Sparse Bayesian Learning Based Channel Estimation for Millimeter-Wave Massive MIMO Systems," in IEEE Transactions on Vehicular Technology, vol. 71, no. 12, pp. 12749-12764, Dec. 2022.
\bibitem{Gong22} Y. Gong, H. Zhao, K. Hu, Q. Lu and Y. Shen, "A Multipath-Aided Localization Method for MIMO-OFDM Systems via Tensor Decomposition," in IEEE Wireless Communications Letters, vol. 11, no. 6, pp. 1225-1228, June 2022. 
\bibitem{Schmidt86} R. Schmidt, "Multiple emitter location and signal parameter estimation," in IEEE Transactions on Antennas and Propagation, vol. 34, no. 3, pp. 276-280, March 1986.
\bibitem{Davies79} D. L. Davies and D. W. Bouldin, "A Cluster Separation Measure," in IEEE Transactions on Pattern Analysis and Machine Intelligence, vol. PAMI-1, no. 2, pp. 224-227, April 1979.
\bibitem{Slepian78} D. Slepian, "Prolate spheroidal wave functions, fourier analysis, and uncertainty — V: the discrete case," in The Bell System Technical Journal, vol. 57, no. 5, pp. 1371-1430, May-June 1978.
USA: Now Publishers, 2006.%
\bibitem{Pati93} Y. C. Pati, R. Rezaiifar and P. S. Krishnaprasad, "Orthogonal matching pursuit: recursive function approximation with applications to wavelet decomposition," Proceedings of 27th Asilomar Conference on Signals, Systems and Computers, Pacific Grove, CA, USA, 1993, pp. 40-44 vol.1.
\end{thebibliography}
\end{document}